\documentclass[aps,pre,twocolumn,superscriptaddress,showpacs,floatfix]{revtex4}

\usepackage{latexsym}
\usepackage{amsmath}
\usepackage{amssymb}
\usepackage{amsfonts}
\usepackage{graphicx}

\begin{document}


\title{Transition to superdiffusive behavior in  intracellular
        actin-based transport \\ mediated by molecular motors}



\author{L.Bruno}

\affiliation{Departamento de F\'{\i}sica, Facultad de Ciencias
Exactas y Naturales,\\ Universidad de Buenos Aires, 1428 Buenos
Aires, Argentina.}
\affiliation{Consejo Nacional de Investigaciones Cient\'{\i}ficas
y T\'{e}cnicas, Argentina.}


\author{V. Levi}

\affiliation{Departamento de Qu\'{\i}mica Biol\'ogica, Facultad de Ciencias
Exactas y Naturales,\\ Universidad de Buenos Aires, 1428 Buenos
Aires, Argentina.}

\affiliation{Consejo Nacional de Investigaciones Cient\'{\i}ficas
y T\'{e}cnicas, Argentina.}


\author{M. Brunstein}

\affiliation{Laboratoire de Photonique et Nanoestructure, Route de Nozay, 91460 Marcoussis, France. }


\author{M. A. Desp\'osito}

\email[]{mad@df.uba.ar}

\affiliation{Departamento de F\'{\i}sica, Facultad de Ciencias
Exactas y Naturales,\\ Universidad de Buenos Aires, 1428 Buenos
Aires, Argentina.}

\affiliation{Consejo Nacional de Investigaciones Cient\'{\i}ficas
y T\'{e}cnicas, Argentina.}


\begin{abstract}

Intracellular transport of large cargoes, such as organelles, vesicles or large proteins,
is a complex dynamical process that involves the interplay of ATP-consuming molecular motors,
cytoskeleton filaments and the viscoelastic cytoplasm. The displacements of particles
or probes in the cell cytoplasm as
a function of time are characterized by different (anomalous) diffusion regimes.
We investigate here the motion of pigment organelles (melanosomes) driven
by myosin-V motors in \emph{Xenopus laevis} melanocytes using a high spatio-temporal resolution
tracking technique.
By analyzing the mean square displacement (MSD) of the obtained trajectories as a
function of the time lag, we show that
the melanosomes display a transition between subdiffusive to superdiffusive behavior.
A stochastic theoretical model is introduced to generalize the interpretation of our data.
Starting from a generalized Langevin equation that explicitly
considers the collective action of the molecular
motors we derive an analytical expression for the MSD as
a function of the time lag, which also takes into account
the experimental noise. By fitting our model
to the experimental data we were able to
discriminate
the exponents that characterize the passive and active contributions to melanosome
dynamics. The model also estimates the ``global" motor forces correctly.
In this sense, our model gives a quantitative description of active transport in living
cells with a reduced number of parameters.

\end{abstract}




\maketitle


\section{Introduction}


Molecular motors are responsible for the active transport of organelles and other
cargoes along cytoskeleton tracks to their correct destination in the cytoplasm.
Three different classes of molecular motors are involved in this task:
dynein and kinesin motors, which transport cargoes toward the minus and plus
ends of microtubules, respectively, and myosin motors, responsible for the
transport along actin filaments toward the barbed end (reviewed in \cite{myo}).
While properties of microtubule-dependent transport, considered to be responsible for
long-distance transport, have been extensively studied \textit{in vivo} \cite{Gross,Levi,BEL,Gross2},
we still do not know important aspects of the properties and regulation of actin-dependent
transport which are believed to support local, short-distance movement of cargoes  in living cells \cite{LangAtk}.

The cytoskeleton (CSK) is a crowded network of
semiflexible linear protein polymers with a complex dynamics, that can exert forces and affect the rheology
of cells \cite{Miz, Bur}.
Particularly, F-actin network has no global
directionality and consists on intercepting randomly distributed filaments \cite{Snider}.
The cytoskeleton, organelles, proteins and motors are part of the highly crowded
 cytoplasm where the  transport of cargoes takes place.


Active (AMR)  and passive (PMR) microrheolgy are two distinct approaches to study the mechanical properties
of the intracellular environment \cite{Wai}. While in AMR  micrometer-sized embedded probe particles are
manipulated by
external fields and their displacement is measured,  the spontaneous displacement
fluctuations of probe particles is analyzed in PMR.

Recent studies using AMR and PMR  have shown that the action of ATP-consuming molecular motors  drives the
system out of equilibrium.  The violation of the fluctuation dissipation theorem (FDT) has been observed in
an \textit{in vitro} system consisting of
a cross-linked actin network with embedded myosin motors \cite{Miz} and also for  beads attached to the cell
membrane of human airway smooth muscle (HASM) cells \cite{Bur}.
Likewise, direct evidence of the deviation from equilibrium inside the cell has been recently obtained
 \cite{Wil,Lau}.
In Ref.\cite{Wil}  both
forced and spontaneous motions of magnetic microbeads engulfed by  \textit{Dyctyostelium} cells
are analyzed to derive the power spectrum of forces acting on intracellular phagosomes.
However, the theoretical model proposed to describe their data did not distinguish between
active and passive contribution to the transport.


In most of PMR experiments it has been observed that the mean square displacement (MSD) of the probes  displays a
crossover between  a subdiffusive  and a superdiffusive regimes \cite{Metz,Bru,Kul,Sal,Gal,Len,Bur}.
The subdiffusive behavior is
characterized by an exponent  ranging from 0.2 \cite{Tre} to $\lesssim 1$  \cite{Tseng,Weiss}
while the  superdiffusive
behavior presents exponents close to 1.5 \cite{Bur,Wil}.
Typically, the transition
time between these two regimes is on the order of 1 s.

While some authors attribute
the subdiffusive behavior to elastic trapping \cite{Tseng}, obstruction \cite{Sax},
crowding \cite{Ban}  or stalling \cite{Bur},
others propose that apparent subdiffusion can arise from
noise  inherent to single particle tracking (SPT) experiments due to  slight errors on the
determination of the actual particle position  \cite{Mart}.
Although there is not a general consensus of the  causes of subdiffusion yet,
 it is well accepted that superdiffusion
has its origin in the collective behavior of molecular motors \cite{Wil,Lau,Bur}.


Properties of motor proteins  have been well characterized \textit{in vitro} but not
in the viscous environment of a cell. Recently, pioneering works
developed new techniques that allow to measure the forces exerted by molecular motors directly \cite{Sht,Shub}.
In Ref. \cite{Sht} motion enhanced differential interference contrast (MEDIC)
movies of living NT2 (neuron-committed teratocarcinoma)
cells at 37°C, was used to determine the force-velocity curves \textit{in vivo}.
In Ref. \cite{Shub} the force that kinesin-1 exerts on lipid droplets
in fly embryos, was measured using a novel development of optical traps that
can be used to apply precise forces to
moving droplets \textit{in vivo}.
Unexpected differences between motor regulation in vivo and in vitro were found, revealing that
further investigation should be done in the field.

\begin{figure}
\begin{center}
\includegraphics[scale=1.8]{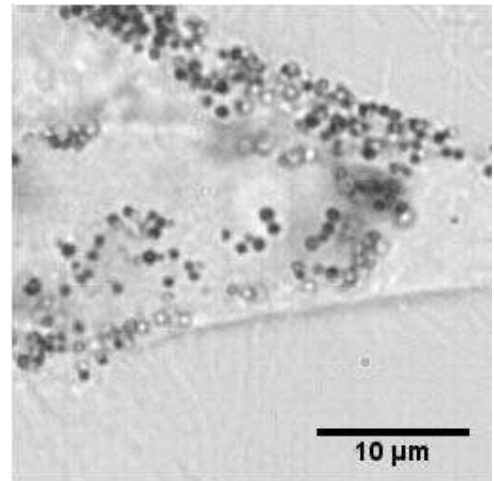}
\end{center}
\caption{A \emph{Xenopus laevis} melanophore image obtained using bright-field transmission light microscopy.
Melanosomes can be clearly identify and followed with high spatio-temporal resolution using a SPT technique.}
\label{figtray}
\end{figure}

Melanophore cells are an exceptionally convenient model system to study intracellular transport
driven by molecular motors \cite{Nasc}, and thus to investigate some properties of
out-of-equilibrium systems in living
cells. Melanocytes have pigment organelles called melanosomes
which are filled with the black pigment melanin. Then, they can be easily imaged using
bright-field transmission light microscopy (Fig.\ref{figtray}) and tracked with millisecond temporal resolution
and nanometer precision \cite{Levi}.
Using a single particle tracking (SPT) technique \cite{Levi3,Gross}, we follow the motion
of myosin-V driven melanosomes along actin filaments and compute the MSD
from the analysis of their trajectories.


%
%


Melanosomes are more or less spherical and stiff particles with sizes around 500 nm.
Pigment organelles can be distributed in the cells in two configurations: either aggregated
in the perinuclear region or homogeneously dispersed in the cytoplasm.
The transport of pigment organelles during aggregation and dispersion is regulated by
signaling mechanisms initiated by the binding of specific hormones to cell surface
receptors \cite{horm}.
In a recent paper, we analyzed the MSD dependence with the time lag using an empirical model \cite{Bru}.
With the aid of numerical simulations, we concluded
that the main difference between aggregation and dispersion condition was the average time the melanosome
spends diffusing during the intervals between successive processive motions. As a consequence, active part of the
transport would not be influenced by stimulation conditions.

In this work we explore an alternative and complementary approach to obtain quantitative information
about the intracellular transport, that can be generalized to other biological data.
Starting from a stochastic model  that  considers the
viscoelasticity of the intracellular environment and the
action of molecular motors explicitly,  we derive an analytical expression for the MSD
as a function of the time lag, which also takes into account the experimental noise.
By fitting the logarithmic derivative of the
MSD-versus-lag time to the experimental data we were able to
 discriminate the passive and active contributions to the melanosome dynamics.
 Experiments using cells expressing a dominant negative construct of myosin-V
where melanosomes are not being actively transport by myosin-V, confirmed the
robustness of the model.
Furthermore, our approach enables us to quantitatively assess the motor forces correlation function,
to estimate the magnitude of motor forces and to determine an effective diffusion coefficient
by only using data coming from a PMR experiment.


\section{Materials and methods}


\subsection{Melanophore cell culture and transfection }

Immortalized \emph{Xenopus laevis} melanophores were cultured as described in Ref.\cite{Rog}.
In order to track the movement of individual organelles, the number of melanosomes
in the cell was reduced by treatment with phenylthiourea \cite{Gross2}.

To study transport along actin filaments, the cells were incubated at 0$^\circ$C for 30
min with 10 $\mu$M nocodazole to depolymerize microtubules \cite{Levi3}.

Melanophores were stimulated for aggregation or dispersion with 10 nM
melatonin or 100 nM MSH, respectively.
The samples were observed between 5 and 15 min after stimulation.
All the measurements were performed at 21$^\circ$C.

Cells were transfected using the FuGENE 6 transfection reagent (Boehringer Mannheim Corp.)
following the vendor's protocols with a plasmid encoding a green fluorescent
protein-tagged myosin-V tail.
Expression of this plasmid results in a dominant-negative inhibition of myosin-V
driven melanosome transport \cite{Rog2}.
The plasmid was a kind gift of Dr. Vladimir Gelfand
(Northwestern University, Chicago, IL)

\subsection{Samples preparation for imaging}

For microscopy measurements, cells were grown for 2 days on 25-mm round
coverslips placed into 35-mm plates in 2.5 ml of the medium.
Before observation, the coverslips were washed in serum-free 70\% L-15
medium and mounted in a custom-made chamber specially designed for the microscope.

\subsection{Tracking experiments}

Single particle tracking experiments of melanosomes moving along actin filaments in wild
type cells were carried out in a Zeiss IM 35 microscope adapted for SPT using a
63x oil-immersion objective (numerical aperture = 1.25) under illumination with a
tungsten-halogen lamp.
A CCD camera (PixelVision, Oregon, US) was attached to the video port of the microscope
for imaging the cells. Movies were registered at a speed of 14 frames/s.

Tracking experiments of melanosomes in transfected cells were carried out in an Olympus IX70
microscope using a 60x water-immersion objective (numerical aperture = 1.2).
A cMOS camera (Pixelink, Ottawa, Ontario, Canada) was attached to the video port
of the microscope for imaging the cells at a speed of 50 frames/s.

Trajectories of melanosomes were recovered with 2 nm precision and 10 ms temporal resolution
from the movies registered as described above using the pattern-recognition algorithm
described in \cite{Levi2}. This algorithm is included in the program Globals for Images developed at
the Laboratory for Fluorescence Dynamics (UCI, Irvine, CA).
The program, which also contains some of the tools used for trajectory analysis,
can be downloaded from the Laboratory for Fluorescence Dynamics website (www.lfd.uci.edu).


\section{Experiments}


Melanophores were treated with nocodazole as described in Materials and Methods
in order to depolymerize microtubules.
After this treatment, aggregation and dispersion of melanosomes were induced by
addition of melatonin and MSH, respectively.
Movies of regions of the cells were recorded from which a total of 134 trajectories
of melanosomes moving along actin filaments were obtained in aggregation and dispersion
by using the pattern-recognition algorithm \cite{Levi2}.


%
\begin{figure}
\begin{center}
\includegraphics[scale=.6]{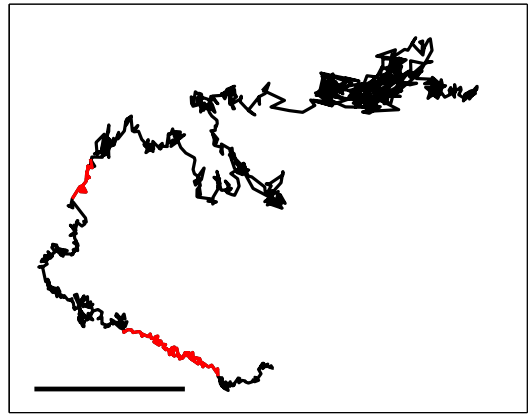}
\end{center}
\caption{ (color online) A representative trajectory of a melanosome followed during 70 s with 0.07 s resolution.
The lighter colored segments represent the periods of rectilinear motion where the speed of the motor-driven motion was computed.
The cell was
stimulated with 100 nM of MSH. Scale bar = 0.5$\mu$m } \label{figtray2}
\end{figure}
%


Figure \ref {figtray2} shows a representative trajectory of a myosin-V driven melanosome during dispersion.
The tracks are noisy; however, periods of rectilinear motion  can be distinguished
from periods of random or diffusive behavior.
Analysis of trajectories during the rectilinear segments gives a rough estimation
of the speeds distribution that can be associated with the motor driven motion.
We obtained a wide distribution for the speed $v$ that can be well described with
a lognormal distribution with mean value $70\pm 17$ nm/s and standard deviation of
$\log v$ equal to $0.6\pm 0.2$ \cite{Bru}.
These values are in agreement with the ones mentioned in the literature for myosin-V \textit{in
 vivo }\cite{Snider}.
The noise of the experimental trajectories, quantified using a smoothing procedure
was between
$1\times 10^{-4}$ and $18\times 10^{-4}$  $\mu m^{2}$, which is much larger than the noise
expected from the tracking
method ($\sim 0.2 \times 10^{-4}$ $\mu m^{2}$) \cite{Mart}.


The mean square
displacement (MSD) for every trajectory is calculated as follows,
\begin{eqnarray}
MSD(\tau)=\left\langle \left(x(t+\tau)-x(t)\right)^2 + \left(y(t+\tau)-y(t)\right)^2 \right\rangle
\label{defmsd}
\end{eqnarray}
where x and y are the coordinates of the particle, $\tau$  is a lag time and the brackets
represents the time average .

%

\begin{figure}
\begin{center}
\includegraphics[scale=.6]{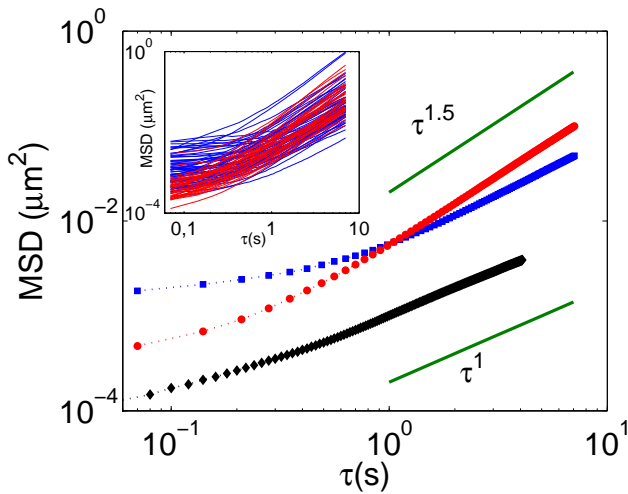}
\end{center}
\caption{(color online) Double-logarithmic plot of the mean square displacement (MSD)
as a function of the time lag. Experimental data points for wild-type cells are
symbolized by squares (aggregation) and circles (dispersion), while the data corresponding
to mutant cells are symbolized by  rhomboids.
The symbols represent an average over all the trajectories.
\textbf{Inset}: MSD vs time lag plot for all  trajectories in wild-type cells. }
\label{figmsd}
\end{figure}

%

Figure \ref{figmsd} shows the average value of MSD obtained during dispersion  and aggregation.
The average distance traveled by the organelles during dispersion is significantly higher
than during aggregation as was previously observed \cite{Snider,Gross}.
 For times larger than 1 second, the behavior presents anomalous diffusion
 with an exponent around 1.35.

The observation of a residual value of the MSD at $\tau \rightarrow 0$ (MSD$_o$)  suggests that
transported melanosomes move in a constrained region with a dynamics
faster than the temporal resolution of the tracking method.
The presence of this jittering also results in an apparent
subdiffusive behavior for short times scales \cite{Mart,Diet2}.

We compute
 MSD$_o$  by linear extrapolation of the first four data points of each experimental MSD vs $\tau$ plot.
The obtained  MSD$_o$ values followed an exponential distribution, with characteristic value equal  to
$13\pm 6\times 10^{-4}$  $\mu m^{2}$
for aggregation and  $2.6\pm 0.1\times 10^{-4}$  $\mu m^{2}$  for dispersion.

It is important to mention at this point that the MSD vs $\tau$ plots for the different trajectories
 present a wide distribution, principally related to
the spreading of the MSD$_o$  values (see the inset in Fig.\ref{figmsd}). This observation was also found in
other systems \cite{Wil,Bur2}. In this sense, the ensamble average plot shown in Fig. \ref{figmsd} represents
 a poor estimator of the behavior of the driven particle.
Instead, the logarithmic derivative of the MSD-versus-lag time $\beta(\tau)$, defined as \cite{Metz}
\begin{eqnarray}
\beta(\tau)=  \frac{d }{d \ln \tau}\ln  MSD(\tau) \, ,
\label{beta}
\end{eqnarray}
redounds on  a more accurate magnitude to describe the dynamics of the transport.
Figure \ref{fittheo} shows $\beta$ as a function of the time lag for the different experimental conditions.


In order to characterize the dynamical properties of melanosomes when they are not
being actively transported by myosin-V, we transfected \textit{Xenopus} melanophores
cells with a plasmid encoding a green fluorescent protein-tagged myosin-V tail.
Since this mutant of myosin-V cannot attach to actin filaments,
expression of this plasmid results in a dominant-negative inhibition of myosin-V
driven melanosome transport \cite{Rog2}.
The MSD$(\tau )$ was
calculated for every trajectory and the average behavior is shown in Fig. \ref{figmsd}.
We also analyzed the value of MSD obtained by extrapolation at  $\tau = 0$ s
for the dominant-negative cells and verified that the MSD$_o$ values obtained  followed a Gaussian
distribution with $0.56 \times 10^{-4}$ $\mu m^{2}$ mean and
$ \sigma  = 0.15\times 10^{-4}$ $\mu m^{2}$. These values are in agreement with the ones expected according
to the error of the tracking method \cite{Mart,Bru}.


\section{Theoretical model}


In this section we apply the quantitative stochastic model
introduced  in an accompanying theoretical paper  \cite{BD}.
We consider the  diffusive behavior of a melanosome of mass $m$,
immersed  in the intracellular medium  and simultaneously
driven by molecular motors.
For this purpose, we describe the resulting dynamics  by means  of  the generalized Langevin equation (GLE)
\begin{eqnarray}
m \ddot{X}(t) + \int_0^t dt' \, \gamma (t-t')\, \dot{X}(t')  =  F(t)    \, , \label{Lang}
\end{eqnarray}
where $\gamma(t)$ is the dissipative memory kernel that characterizes the viscoelastic
properties of the cytoplasm. The random force $F(t)$ is assumed to be the sum of two uncorrelated
 contributions
\begin{eqnarray}
F(t)=\xi(t)+\chi(t)
 \, \label{fr}
\end{eqnarray}
being  $ \xi(t) $ the \textit{internal} noise due to thermal activity,
while $\chi(t)$  is an \textit{external} random force representing the action of the molecular motors.
%


The thermal  noise $ \xi(t) $ is a zero-centered and stationary
random force with correlation function
\begin{eqnarray}
\langle \xi(t) \xi(t') \rangle = C(|t - t'|)
 \, ,\label{rint}
\end{eqnarray}
and is related  to the memory kernel  $\gamma(t)$   via the fluctuation-dissipation
theorem \cite{Zwa}
\begin{eqnarray}
C(t)= k_B T \, \gamma(t)
 \, ,\label{tfd}
\end{eqnarray}
where $T$ is the absolute temperature, and $ k_B $ is the Boltzmann
constant.
This internal noise is the responsible for the passive
transportation.
To reproduce the subdiffusive behavior due to the viscoelastic properties of the cytoplasm,
the noise autocorrelation
function (\ref{rint}) can be modeled as  a power-law \cite{powerlaw}
\begin{eqnarray}
C(t)=  \frac{C_0}{\Gamma(1-\lambda)}\,\left(\frac{t}{\tau_0}\right)^{-\lambda} \, ,
\label{nint}
\end{eqnarray}
where $0<\lambda <1 $,  $C_0$ is a proportionality coefficient,
$\tau_0$ is set to 1 s., and $\Gamma(z) $ is the Gamma function.


On the other hand, the random force  $\chi(t)$ is not related
to the memory kernel $\gamma(t)$
 because it stems from the activity of the  molecular motors.
The irreversible conversion of chemical energy  from ATP hydrolysis into the particle motion \textit{via}
the activity of  myosin-V motors drive the system to an out-of-equilibrium situation.
As a consequence,
the  FDT   is no longer valid.

Assuming   that the actin network  has no global directionality,
the random force $\chi(t)$
exerted by the action of the myosin-V motor is chosen as a zero-centered one.
Moreover, in recent experiments it was observed that the forces power spectrum
exhibits a power law behavior \cite{Lau,Bur,Wil,Gal}.
Then,  we assume that the autocorrelation function
\begin{eqnarray}
\langle \chi(t) \chi(t') \rangle = \Lambda(|t - t'|)
 \, \label{rext}
\end{eqnarray}
can be written as \cite{BD}
\begin{eqnarray}
\Lambda(t)=  \frac{\Lambda_0 } {\Gamma(1-\alpha)}\,\left(\frac{t}{\tau_0}\right)^{-\alpha} \,
, \label{corract}
\end{eqnarray}
where $0<\alpha <1$, $\Lambda_0$ is a proportionality coefficient and $\tau_0$ is set to 1 s.


Considering that the average melanosome diameter is 500 nm \cite{Rog},
its density is $\sim$ 1.2 g/ml \cite{Sha} and $\gamma \sim 10^{-5}$ Ns/m
(see Refs. in \cite{Yam}), the characteristic time  would be $\sim$ $10^{-12}$ s. Then,
from an experimental point of view, the inertial effects can be neglected.
In this situation,  the temporal evolution of the displacement
can be obtained from (\ref{Lang}), which is given by
\begin{eqnarray}
X(t) &= & x_0 +  \int_0^t dt' G(t-t') (\xi(t') +
\chi(t')) \, ,
\label{X}
\end{eqnarray}
where $x_0=X(t=0)$ is the deterministic
initial position  of the particle.
The  relaxation function  $G(t)$ is the inverse form of the
Laplace transform
\begin{eqnarray}
\widehat{ G}(s) =  \frac{1}{ s \widehat{\gamma}(s) } \, ,
 \label{kernella}
\end{eqnarray}
where $\widehat{\gamma}(s)$ is the Laplace transform of the
dissipative memory kernel.


To obtain an analytical expression of the MSD it is necessary to calculate the two-time position correlation.
Starting from Eq. (\ref{X}) and using relation (\ref{tfd}), it
 can be written as
\begin{eqnarray}
&&\langle X(t+\tau)X(t) \rangle  =
  x_0^{2} +
 k_B T \, \left(I(t) + I(t+\tau) - I(\tau)\right) \nonumber \\
&& \qquad +  \int_0^{t} dt_1
 \left( G(t_1) H(t_1+\tau)+ G(t_1+\tau) H(t_1) \right)
\, ,\nonumber \\
\label{xx}
\end{eqnarray}
%
%
%
where the involved relaxation functions are given by
\begin{eqnarray}
 I(t) &=&  \int_0^t dt' G(t')\, ,
 \label{relaIG}
\\
H(t) &=&  \int_0^t dt' G(t') \Lambda(t-t')  \, . \label{H}
\end{eqnarray}
$I(t)$ only depends on the internal noise, while $H(t)$ is related to both internal
and external contributions.


Using the   autocorrelation functions (\ref{nint}) and (\ref{corract}),
 the involved relaxation functions can be written as
\begin{eqnarray}
I(t)  & = & \frac{k_B T}{C_0}\frac{1}{\Gamma(\lambda+1)} \,\left(\frac{t}{\tau_0}\right)^{\lambda} \, , \\
G(t)  & = & \frac{k_B T}{\tau_0 \, C_0 }\frac{1}{\Gamma(\lambda)} \,\left(\frac{t}{\tau_0}\right)^{\lambda-1} \, , \\
H(t) & = & \varepsilon \, k_B T \,  \frac{1}{ \Gamma(\lambda-\alpha+1)} \,\left(\frac{t}{\tau_0}\right)^{\lambda-\alpha}\,   ,
\end{eqnarray}
where
\begin{eqnarray}
\varepsilon = \frac{\Lambda_0}{C_0}
\label{varepsilon}
\end{eqnarray}
is a dimensionless parameter that measures the relative intensity among
the motors forces and the thermal  forces.


For $2 \lambda -\alpha>0$  the  integral term of (\ref{xx}) can be explicitly evaluated.
In this case,  it can be demonstrated that the MSD long  time limit
\begin{eqnarray}
MSD(\tau)=  \lim_{ t \to \infty } \langle \left(X(t+\tau)-X(t)\right)^{2} \rangle
\label{ltlmsd}
\end{eqnarray}
 can be written as \cite{BD}
\begin{eqnarray}
MSD(\tau) &= &
\frac{4k_B T}{\gamma_0}
\left\{\frac{1}{\Gamma (\lambda +1)}(\frac{\tau}{\tau_0})^{\lambda }
+ \varepsilon  K_{\lambda,\alpha}(\frac{\tau}{\tau_0})^{2 \lambda -\alpha }\right\}  \, ,
\nonumber \\
\label{msdanomal}
\end{eqnarray}
where
\begin{eqnarray}
K_{\lambda,\alpha}=\Gamma (\alpha -2 \lambda ) \, \frac{\sin (\pi  (\lambda -\alpha ))-\sin (\pi  \lambda )}{\pi }
\, ,
\end{eqnarray}
is a positive constant and  $\gamma_0 = C_0/kT$. The two-dimensional situation is considered by
the factor 4.

It is worth mentioning that, although $0< \alpha <1$,
when  $1<2 \lambda -\alpha <2$ the MSD (\ref{msdanomal}) exhibits a crossover from a
subdiffusive regime  with an exponent $\lambda$,  to a superdiffusive
regime with an exponent $2 \lambda -\alpha$ \cite{BD}.
The first term of (\ref{msdanomal}) represents  the passive transport of the melanosome
in the viscoelastic medium, while
the second one corresponds to  the contribution of the random force
$\chi(t)$ and is originated in the activity of the myosin-V motors.


Finally, to compare the analytical expression (\ref{msdanomal}) with the  experimental data
obtained using SPT techniques, it is necessary to take into
account the error on the particle position determination.
As is established in Refs.\cite{Mart,Diet}, the effect of an uncorrelated noise of variance
$\eta^2$ -generated by measurement errors in particle location or by biological activity-
is to add a constant to
the mean square displacement.
Then, the $MSD(\tau)$ given in
(\ref{msdanomal}) rewrites as
\begin{eqnarray}
MSD(\tau)\rightarrow MSD(\tau) + (2\eta)^2 \, .
\label{msdexp}
\end{eqnarray}

As it was previously discussed, the corresponding  local slope
of the MSD-versus-lag time (\ref{beta}) is a better function to characterize the dynamics.
From (\ref{msdanomal}) and (\ref{msdexp}) we get
\begin{eqnarray}
\beta(\tau) &= &
\frac{ \frac{\lambda}{\Gamma (\lambda +1)}(\frac{\tau}{\tau_0})^{\lambda }+\varepsilon \,
(2 \lambda -\alpha)\, K_{\lambda,\alpha}\,(\frac{\tau}{\tau_0})^{2 \lambda -\alpha }}
{\frac{1}{\Gamma (\lambda +1)}(\frac{\tau}{\tau_0})^{\lambda}+\varepsilon
\,  K_{\lambda,\alpha}\,(\frac{\tau}{\tau_0})^{2 \lambda -\alpha }+ \delta}
\label{betaexp} \, ,
\nonumber \\
\end{eqnarray}
where
\begin{eqnarray}
\delta =\gamma_0 \frac{(2\eta)^2}{4 k_B T}
 \, \, . \label{delta}
\end{eqnarray}

In particular, setting  $\varepsilon=0$ in  Eq. (\ref{betaexp}) one gets
 \begin{eqnarray}
\beta(\tau) &= &   \, \frac{\lambda \, (\frac{\tau}{\tau_0})^{\lambda }}
{(\frac{\tau}{\tau_0})^{\lambda }+ \Gamma (\lambda +1)\, \delta }
\label{betamut}  \, .
\end{eqnarray}
which corresponds to the absence of active transport case.


\section{Results and Discussion}


In this paper we study the motion of 500 nm melanosomes in \emph{Xenopus laevis} melanophores
treated with nocodazole (a microtubule depolymerizer), where the active transport is
powered by the actin-dependent motor myosin-V. Aggregation and dispersion of melanosomes
were induced by the addition of  melatonin and MSH, respectively.

\begin{figure}
\begin{center}
\includegraphics[scale=.6]{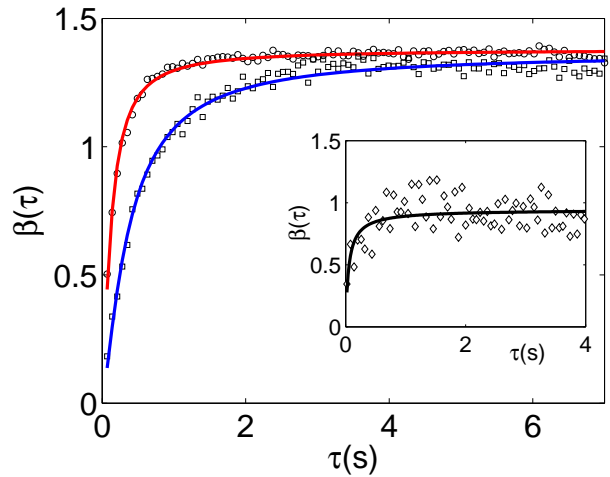}
\end{center}
\caption{The logarithmic derivative of the mean square displacement as a function of the
time lag (Eq. \ref{beta}). The experimental data for wild-type cells are symbolized with squares (aggregation)
and circles (dispersion). The solid lines represent the fit to $\beta$ of the expression given by
Eq. (\ref{betaexp}). \textbf{Inset}: The same plot for mutant cells. The solid line represents the fit of
Eq. (\ref{betamut}).
The symbols represent ensemble averages.}
\label{fittheo}
\end{figure}

The MSD of melanosomes in both stimulation conditions displayed a transition from a
subdiffusive to a superdiffusive regime and the corresponding local slope of the
MSD-versus-lag time was well described by Eq. (\ref{betaexp}), as shown in Fig.\ref{fittheo}.

Four parameters characterize the behavior: $\lambda$, $\alpha$, $\varepsilon$ and $\delta$,
where $\lambda$ and $\alpha$ are the power law exponents of the internal and external noise
correlation functions, $\varepsilon$ is a parameter that measures the relative intensity
between random forces and $\delta$ is associated with the residual value of the
MSD as $\tau\rightarrow 0$.


%



\begin{table}[h!]
  \centering

  \begin{tabular}{|c|c|c|c|c|}
    \hline
     & $\lambda$ & $\alpha$ &  $\varepsilon$ &   $\delta$  \\
     \hline
    aggregation &  $0.96 \pm 0.04 $ & $0.58\pm 0.08$ & $202 \pm 38$  & $41 \pm 8$ \\
    \hline
    dispersion  & $0.98 \pm 0.03$ & $0.58 \pm 0.10$   & $83 \pm 13$    & $4.0 \pm 0.6 $  \\
    \hline
    mutant m-V  & $0.94 \pm 0.04$ & -   & -    & $0.061 \pm 0.014 $  \\
    \hline
  \end{tabular}
 \caption{Dimensionless parameters of the model $\lambda, \alpha, \varepsilon$ and $\delta$ were obtained from
the fit of Eqs. \ref{betaexp} and \ref{betamut} to the experimental data shown in Fig. \ref{fittheo}. The values
are given as mean $\pm$ standard deviation of the ensemble.}
  \label{table1}
\end{table}


The parameters $\lambda$ , $\alpha$ , $\varepsilon$ and $\delta$ obtained from the fitting shown in
Fig.\ref{fittheo}  are displayed in Table \ref{table1}.
The power-law exponents $\lambda$ and $\alpha$ for both dispersing and aggregating melanosomes
were similar ($\lambda\sim 0.97$ and $\alpha\sim 0.58$).
These values lead to an asymptotic exponent
$2\lambda-\alpha \sim 1.36$, which is a signature of superdiffusion.
Similar values were obtained in other systems \cite{Tre,Wil,Gal,Bur}.
On the other hand, the obtained value $\alpha\sim 0.58$
implies a power spectrum of fluctuating motor forces that scales as $\omega^{-0.42}$,
corresponding
to a smoothing of discontinuities in instantaneous
force pulses, as was stated in Refs. \cite{Wil,Bur}.


It is worth pointing out that with our model  we are able to discriminate the
contributions of internal and external forces  in a straightforward fashion.
In a recent paper \cite{Wil}, Wilhelm described the dynamics of forced motions of magnetic microbeads
in \textit{Dictyostelium} cells  in terms of a generalized
Langevin equation.
However, differently of what we have done in this paper, that model considers a single force term
which included contributions of both thermal  Brownian forces and driving forces
generated by molecular motors.
The force correlation function followed a power-law behavior with exponent 0.8,
which is between the values of $\lambda$ and $\alpha$  obtained by us.


Melanosomes in cells expressing a dominant-negative myosin-V construct,
and thus not being actively transported by myosin-V, show spontaneous motion
and the MSD displays   a subdiffusive   behavior.
In the absence of motors,  Eq. (\ref{betaexp}) rewrites in terms of $\lambda$ and
$\delta$ (Eq. (\ref{betamut})).
Interestingly, we obtained a value for $\lambda \sim 0.94$, very similar to the
one obtained in the presence of active motors. This reinforces our hypothesis
 that $\lambda$ is the parameter that characterizes the subdiffusive motion of the melanosomes
 in the  cytoplasm.
Therefore, our model allows us to discriminate
between passive and active contributions to the motion.
In both cases (i.e. in the presence and in the absence of active motors),
$\lambda$ is close to 1. Similar values were obtained for the subdiffusive anomalous exponent
of tracer proteins in highly concentrated random-coil polymer and globular protein
solutions that mimic the crowded conditions encountered in cellular environments
\cite{Ban}, and for the anomalous diffusion of dextran
polymers inside HeLa cells \cite{Weiss}.


Typically, the anomalous diffusion regime is characterized by fitting an empirical power-law
relationship to the MSD dependence with $\tau$  of the form \cite{Bru,Bur2,Rau}:
\begin{eqnarray}
MSD(\tau) &= & C+D^{*} (\frac{\tau}{\tau_0})^{ \gamma}  \, ,
\label{empirical}
\end{eqnarray}
being $D^{*}$  an effective diffusion coefficient and $\gamma$ an exponent between 0 and 2.
Notice that with this definition $D^{*}$ has units of distance square.

Likewise, we can define the effective diffusion coefficient for a subdiffusive and
a superdiffusive limit regimes in our model. In the first case, setting $\varepsilon=0$
and $\gamma=\lambda$ in (\ref{msdanomal}), we have that
 \begin{eqnarray}
 D^{*}_{sub} &= &
\frac{(2\eta)^2}{\delta}
\frac{1}{\Gamma (\lambda +1)}
\label{Dsub}
\end{eqnarray}

In the superdiffusive limit, setting $\gamma=2 \lambda -\alpha$ in Eq. (\ref{msdanomal})  we obtain
\begin{eqnarray}
 D^{*}_{sup}=  \varepsilon\,\frac{(2\eta)^2}{\delta}  K_{\lambda,\alpha} \, ,
\label{Dsup}
\end{eqnarray}
where we have used relation
(\ref{delta}) to derive this expression.

Interestingly, these analytical expressions enable a quantitative interpretation of the physics underlying
the transport processes, linking the macroscopic effective diffusion coefficient with microscopic parameters
derived from the forces correlation functions.

In Ref. \cite{Bru} we obtained the values of $D^{*}$ by fitting the MSD-vs-time lag plot of each trajectory
with the empirical expression
 given by Eq. (\ref{empirical}). The distributions of $D^{*}$  followed an exponential behavior, with
mean values equal to  $2  \times 10^{-3} \mu m^{2}$ and  $4.4  \times 10^{-3} \mu m^{2}$ for aggregation and dispersion
conditions, respectively.

Using (\ref{Dsup})  and the parameters displayed in Table 1 to compute the
analytical effective diffusion coefficient, we obtained $ (5 \pm 2)\times 10^{-3} \mu m^{2}$
and $(4 \pm 1) \times 10^{-3} \mu m^{2}$, for aggregation and dispersion, respectively. These values are consistent
with the ones obtained empirically.

For mutant cells, the computed analytical effective diffusion coefficient (\ref{Dsub}) is
 $D^{*} = (0.85 \pm 0.1) \times 10^{-3} \mu m^{2}  $, which is in agreement with the one obtained in Ref. \cite{Bru}
($ \sim 1 \times 10^{-3} \mu m^{2} $)
 %


Moreover, the correlation function (\ref{corract}) allows obtaining an estimation
for the magnitude of the ``global" force
exerted by the motors $F_{mot}$, which is given by
\begin{eqnarray}
F_{mot} \approx \sqrt{\frac{\Lambda_0 } {\Gamma(1-\alpha)} }    \, \, .
\label{motfor}
\end{eqnarray}

Using definitions  (\ref{varepsilon}) and (\ref{delta}),  $\Lambda_0$ rewrites as
\begin{eqnarray}
\Lambda_0 = \varepsilon  \frac{\delta }{(2\eta)^2}(2 k_B T)^2 \,  ,
\end{eqnarray}
Taking $k_B T =4 \times10^{-3} pN \mu m$, $(2\eta)^2=$MSD$_o$ and $\varepsilon$, $\delta$ and $\alpha$ from Table
I, we obtain $F_{mot} \sim 8 \pm 4 pN$  and $\sim   16 \pm 6 pN$, for dispersion and aggregation, respectively, which
are in the order of the reported stall force of a single myosin-V \cite{stall}.
The values obtained for aggregation and dispersion are similar within the experimental error,
since as far as we know the stall forces for myosin in these stimulation conditions are expected
to be the same.



\section{Summary}


The violation of the FDT in living cells has been observed in recent studies \cite{Bur,Lau,Wil,Gal}.
Motor proteins, which are force generators in cells, can not only
modify the viscoelastic response of the cytoplasm -as is the case of CSK motors-,
but they also generate non-thermal random forces which
drive the system out of equilibrium \cite{Lau}.
The action of motors is generally reflected in a transition between a subdiffusive to a superdiffusive
behavior of the displacement of particles within the cells \cite{Bur}.

To investigate how the action of molecular motors affects the transport in living cells,
we study the motion of melanosomes driven by myosin-V along the F-actin network in
\emph{Xenopus laevis} melanocytes.

Differently to other microrheology experiments reported before \cite{Bur,Lau,Wil,Gal, Miz},
we adopt a different approach which uses a generalized Langevin equation that explicitly
considers the collective action of the molecular motors and the presence of experimental/biological noise. The
analytical solution of the model is obtained and an expression for the slope of the MSD vs the time lag is
derived explicitly. The comparison between the model solution and the experimental data is straightforward,
and involves 4 parameters. Two of these parameters are  the exponents of the thermal and non-thermal
forces correlation functions, while the others are related with the relative intensity between both kind of forces
and the  residual value of the MSD.

On one hand, the model predicts the observed crossover between subdiffusive to superdiffusive regimes,
as well as it gives
good estimates for the {\it in vivo} motor forces. On the other hand,  the proposed method
enables us to determine a link between the macroscopic effective diffusion
coefficient and the parameters in the microscopic scale.

We  believe that this theoretical approach can be used to describe the dynamics of
intracellular transport of different cargoes in other living cells.

%
\begin{acknowledgments}

We acknowledge
support from grants PICT 928/06, PICT 31980/05 and PICT 31975/05 from Agencia Nacional de Promoci\'{o}n Cient\'{i}fica y
Tecnol\'{o}gica, Argentina.

\end{acknowledgments}



\end{document}